%% file: main.tex
\PassOptionsToPackage{table}{xcolor}
\documentclass[sigplan,nonacm]{acmart}
\usepackage{xspace}
\usepackage{cleveref}
\usepackage{enumitem}
\usepackage{xurl}
\usepackage{subcaption}
\usepackage{pgfplots}
\pgfplotsset{compat=1.18}
\usepgfplotslibrary{groupplots}
\usepackage{algorithm}
\usepackage{algpseudocode}

\newcommand{\ie}{\textit{i.e\@.}}

\AtBeginDocument{%
  }

\acmISBN{978-1-4503-XXXX-X/2018/06}

\newcommand{\sys}{OasisKV\xspace}

\usepackage{pifont}
\newcommand{\circled}[1]{\ding{\the\numexpr181+#1\relax}}

\crefname{figure}{fig.}{figs.}
\Crefname{figure}{Fig.}{Figs.}
\crefformat{section}{\S#2#1#3}
\crefmultiformat{section}{\S#2#1#3}{ and \S#2#1#3}{, \S#2#1#3}{, and\S#2#1#3}
\Crefformat{section}{\S#2#1#3}
\crefformat{subsection}{\S#2#1#3}
\Crefformat{subsection}{\S#2#1#3}
\crefmultiformat{subsection}{\S#2#1#3}{ and \S#2#1#3}{, \S#2#1#3}{, and\S#2#1#3}
\crefformat{subsubsection}{\S#2#1#3}
\Crefformat{subsubsection}{\S#2#1#3}
\crefmultiformat{subsubsection}{\S#2#1#3}{ and \S#2#1#3}{, \S#2#1#3}{, and\S#2#1#3}

\begin{document}

\title[]{OasisKV: Scaling In-Decode KV Cache Beyond HBM with Lookahead Sparse Prefetching}

\author{Can Xiao}
\authornote{Both authors contributed equally to this work. Part of this work was completed during their internships at Microsoft Research.}
\affiliation{%
  \institution{Imperial College London}
  \country{}
  }
\email{cx922@ic.ac.uk}

\author{Sukmin Cho}
\authornotemark[1]
\affiliation{%
  \institution{KAIST}
  \country{}
  }
\email{smcho@casys.kaist.ac.kr}

\author{Junbong We}
\affiliation{%
  \institution{KAIST}
  \country{}
  }
\email{jbwe@casys.kaist.ac.kr}

\author{Zhixiong Niu}
\affiliation{%
  \institution{Microsoft Research}
  \country{}
  }
\email{zhixiong.niu@microsoft.com}

\author{Jianyi Cheng}
\affiliation{%
  \institution{University of Edinburgh}
  \country{}
  }
\email{jianyi.cheng@ed.ac.uk}
  
\author{Yiren Zhao}
\affiliation{%
  \institution{Imperial College London}
  \country{}
  }
\email{a.zhao@imperial.ac.uk}

\author{Youngjin Kwon}
\affiliation{%
  \institution{KAIST}
  \country{}
  }
\email{yjkwon@casys.kaist.ac.kr}

\author{Yongqiang Xiong}
\affiliation{%
  \institution{Microsoft Research}
  \country{}
  }
\email{yongqiang.xiong@microsoft.com}

\author{Rui Ma}
\authornote{Corresponding authors.}
\affiliation{%
  \institution{Microsoft Research}
  \country{}
  }
\email{mrui@microsoft.com}

\author{Junyi Liu}
\authornotemark[2]
\affiliation{%
  \institution{Microsoft Research}
  \country{}
  }
\email{junyi.liu@microsoft.com}

\begin{abstract}
    Large language model (LLM) inference serving is increasingly constrained by memory rather than compute. As long-context and long-form reasoning workloads become more prevalent, the key-value (KV) cache dominates both memory footprint and memory traffic at the token generation (\ie, decode) stage. 
    HBM capacity has therefore become a critical bottleneck, constraining inference batch size and throughput. This paper introduces \emph{\sys}, a memory-centric LLM inference system that alleviates HBM limitations by offloading the KV-cache to alternative memory tiers. Exploiting the sparsity of decode-time attention and leveraging speculative decoding to predict important tokens, \sys selectively prefetches and stages only the most relevant KV entries in HBM for efficient LLM inference computation.
    To support growing LLM context and the larger batches enabled by sparse HBM residency, \sys{} employs tiered KV management across decode, prefill, and remote-memory nodes, relaxing decode-host memory capacity limits under disaggregated serving. 
    An asynchronous prefetch pipeline hides off-GPU transfer latency while bounding per-step traffic to fit the available bandwidth.

  We implement \sys{} on vLLM. With only 2{,}048 KV tokens resident in HBM, \sys{} maintains average accuracy within a negligible 0.61 points of full attention on LongBench~v2 for long-context understanding and across AIME24--25 and GPQA-Diamond for long-form reasoning. This sparsity translates into a $1.86\times$ throughput gain over vLLM on AIME24.
  Synthetic long-context workloads further show up to $2.2\times$ higher throughput in multi-GPU serving. Under prefill--decode disaggregation, \sys{} achieves $2.29\times$ vLLM's throughput while supporting $2.42\times$ as many concurrent requests under the same host-memory budget.
    
\end{abstract}

\maketitle

\input{tex/01_introduction}

\input{tex/02_background_motivation}

\input{tex/03_design}

\input{tex/04_evaluation}

\input{tex/05_related_work}

\input{tex/06_conclusion}

\newpage

\bibliographystyle{ACM-Reference-Format}
\bibliography{main}

\clearpage
\appendix

\input{tex/07_appendix_compare_lserve}
\newpage
\input{tex/08_appendix_fetch_cap_longbench}
\newpage

\input{tex/09_appendix_compare_spec}
\newpage

\input{tex/10_appendix_combine_spec}
\newpage
\input{tex/11_appendix_alpha}
\newpage
\input{tex/12_appendix_kv_prediction}
\newpage
\input{tex/13_appendix_pd_250gib}

\end{document}

%% file: tex/01_introduction.tex
\section{Introduction}

\begin{figure}[t]
\centering
\includegraphics[width=\columnwidth]{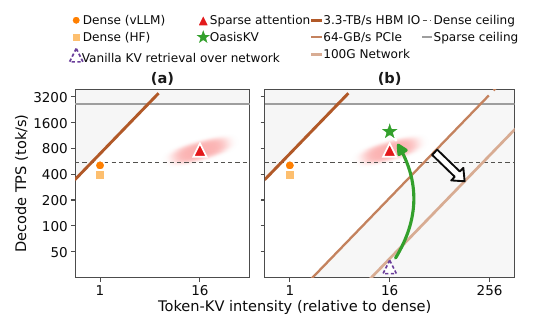}
\caption{KV-cache memory roofline for Qwen3-8B with a 32K-token context.
Token-KV intensity (tokens/byte) is the number of tokens generated per KV-cache byte transferred over the corresponding interface.
The x-axis is normalized to dense attention ($1 \times$).
The red triangular sparse-attention point is LServe~\citep{yang2025lserve}.
At the same HBM-capacity-limited batch size, the dense ceiling represents the maximum throughput with full KV reads, while the sparse ceiling represents the ideal limit with no KV reads.
}
\label{fig:mem_roofline}
\end{figure}

Agentic workloads on web-use, computer-use, and coding tasks are naturally multi-turn and can consume over $10\times$ the context of chatbot-era single-turn requests \citep{bfcl,osworld,Deng2025SWEBenchPC,liu2026agenticcodingwildcharacterizing,zhao2026heterogeneous}. 
As the context length grows during autoregressive decoding, both HBM capacity and bandwidth become critical limitations.
The \emph{capacity limitation} bounds the inference batch size because each request must retain its accumulated KV cache in HBM.
The \emph{bandwidth limitation} constrains decoding speed since dense attention requires frequent KV cache reads.
Improving decode throughput therefore requires addressing both limitations together.

\emph{Sparse attention} reduces KV reads by focusing only on a selected subset of previous tokens when generating new tokens \citep{streamingllm,minference,quest,flexprefill,dsa,seerattention,yang2025lserve}.
\Cref{fig:mem_roofline}(a) illustrates this benefit using a KV-cache memory roofline, which plots measured decode throughputs against Token-KV intensity (\ie, the number of output tokens generated per byte of KV data moved across the corresponding memory interface).
By reducing HBM reads per output token, sparse attention increases Token-KV intensity and lowers time per output token (TPOT), improving throughput (TPS) as shown by the red region in \Cref{fig:mem_roofline}(a).
However, it does not reduce the HBM capacity required to retain the full KV cache, leaving the HBM capacity limitation on batch size unchanged.

Overcoming HBM capacity limitations necessitates offloading the full cache out of HBM and retaining only an actively used subset there. 
However, this approach often incurs a substantial performance penalty, as system throughput then becomes constrained by the slowest memory interface.
This challenge is exacerbated by the rise of disaggregated deployments in modern serving systems, where KV generation, storage, and consumption are distributed across multiple nodes \citep{splitwise,distserve,mooncake,dualpath}. 
While such architectures provide access to a larger collective memory pool for KV storage, retrieving KV data over lower network bandwidth across data-center network hierarchy further constrains the effective I/O bandwidth ceiling (hollow triangle in \Cref{fig:mem_roofline}(b)). 
This observation motivates our core research question: \textit{How can we exploit sparse prefetching to overcome bandwidth bottlenecks and fully benefit from increased KV cache capacity in disaggregated environments?}

Prior work has explored sparse KV retrieval, but these approaches face two key limitations: 
(i) they primarily focus on CPU offloading that is still bounded by precious host DRAM availability, and (ii) their inefficient implementation of compute kernels and inference stacks obscures the underlying bandwidth bottlenecks. 
\emph{KV retrieval} \citep{chen2024arkvale,kvdrive,chen2026retroinfer,spin,hisparse,shadowkv} increase effective KV-cache capacity by storing the complete KV-cache in host memory and staging only the \textit{active} subset in GPU HBM. 
As a result, on-demand retrieval of nonresident KV blocks from host memory adds an extra step to the decoding pipeline.
To address this overhead, \emph{KV prefetch} techniques \citep{specache,yang2025attentionpredictor,infinigen,specontext,sparda,freekv} predict important KV blocks and speculatively prefetch them, thereby overlapping memory transfer latency with decoding computation. 
However, the effectiveness of prefetching critically depends on the length of the compute pipeline --- it must be long enough to hide data transfer latencies. 
Most existing solutions are prototyped in the HuggingFace (HF) Transformers library \citep{hf-transformers}, whose inefficient pipeline artificially allows excessive slack for prefetching, making these strategies less applicable to optimized, real-world serving environments.

\begin{figure}[t]
\centering
\includegraphics[width=\columnwidth]{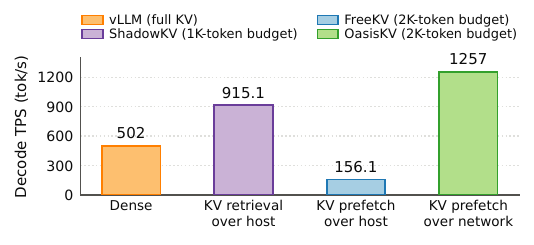}
\caption{Decode throughput of representative KV retrieval~\citep{shadowkv} and KV prefetch systems~\citep{freekv}, and of
\sys, against a dense vLLM baseline on one H100 at a 32K context.
Token budget denotes KV subset size used in attention.}
\label{fig:intro_bars}
\end{figure}

These limitations call for accurate, low-overhead KV prediction and a non-blocking prefetch pipeline that can fit in modern serving environments.
We present \sys, a memory-centric LLM serving system that addresses these requirements through three key design features: lookahead attention, tiered KV placement, and non-blocking asynchronous prefetching, enabling decoding throughput to exceed the dense ceiling and approach the sparse ceiling (\Cref{fig:mem_roofline}).
Even when fetching KV from multiple remote nodes, \sys achieves significantly higher decoding throughput than prior single-node KV retrieval and prefetching systems and dense vLLM (\Cref{fig:intro_bars}).

This paper makes the following contributions:
\begin{itemize}[leftmargin=*]
    \item We introduce lookahead attention that reuses existing speculative draft tokens~\citep{eagle3} to accurately predict per-head KV demand one step ahead with low overhead, while remaining compatible with speculative decoding.
    \item We expand KV-cache capacity beyond GPU HBM and decode-host DRAM through tiered KV placement, using decode-host DRAM as an intermediate cache between remote memory and GPU HBM to reduce redundant remote reads and benefit from nearly infinite DRAM across remote nodes.
    \item We mitigate off-GPU bandwidth limitations with non-blocking asynchronous prefetch. \sys hides latency through fine-grained data movement stages and manages prefetch traffic through hierarchical LRU-based eviction strategy, preventing off-GPU bandwidth from becoming system performance bottlenecks.
    \item We implement \sys on vLLM and evaluate it across dense and MoE models in single-GPU, multi-GPU, and PD-disaggregated deployments. Compared with vanilla vLLM, \sys improves decode throughput by $1.86\times$ on a reasoning workload, by up to $2.2\times$ on multi-GPU long-context serving, and by $1.4$--$2.3\times$ under PD disaggregation, all within a negligible 0.61 points of full-attention accuracy on average.
\end{itemize}

%% file: tex/02_background_motivation.tex
\section{Background and Motivation}
\label{sec:background}
\label{sec:motivation}

This section firstly introduces sparse attention, KV retrieval, and KV prefetch. It then highlights the remaining prediction, capacity, and bandwidth challenges tackled by \sys.

\subsection{Sparse Attention and KV Offloading}
\label{sec:background-serving}
\label{sec:background-sparse-prefetch}

\begin{figure}[t]
	\centering
	\includegraphics[width=1.0\linewidth]{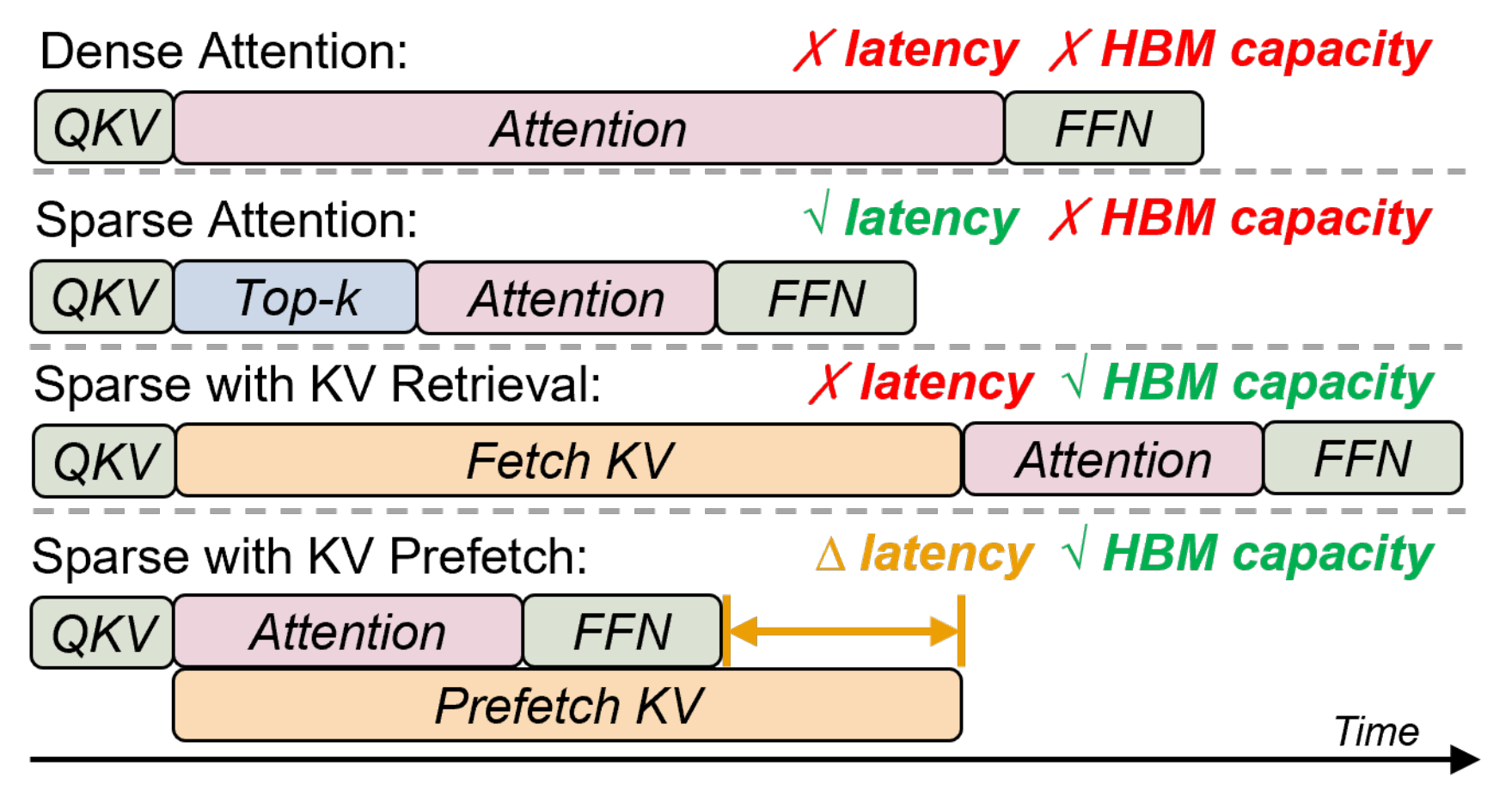}
	\caption{Per-layer execution flows of dense attention, sparse attention, KV retrieval, and KV prefetch. KV retrieval and KV prefetch all fetch from host DRAM.}
	\label{fig:background-flow-comparison}
\end{figure}

To understand the memory cost of long-context decoding, we first characterize how the KV cache grows and is accessed. At each decode step, a transformer appends the newly projected key and value to the KV cache and attends over all preceding entries. Prefill materializes the prompt KV cache, which decode then grows token by token throughout generation. For a transformer with $N_{\mathrm{layer}}$ layers, $N_{\mathrm{kv}}$ KV heads, head dimension $d_h$, and $e$ bytes per KV element, the per-token and per-batch KV footprints are
\begin{equation}
s_{\mathrm{KV}} = 2N_{\mathrm{layer}}N_{\mathrm{kv}}d_he,
\qquad
M_{\mathrm{KV}}(N,L) = NLs_{\mathrm{KV}},
\label{eq}
\end{equation}
where the factor of two accounts for keys and values, and $N$, $L$ represent batch size and sequence length respectively. 

\Cref{fig:background-flow-comparison} illustrates how sparse attention, KV retrieval, and KV prefetch progressively reduce HBM traffic and residency.
Under dense attention, each layer reads the complete accumulated KV history from HBM.
Sparse attention first reduces the per-layer KV traffic by selecting only a subset of the cached blocks. 
It inserts top-$K$ selection between query projection and attention: the current query $q_l^t$ ranks lightweight KV metadata and produces a per-head block set $S_l^t$, after which attention reads only the selected blocks. 
The entire KV cache, however, remains resident in HBM. KV retrieval further reduces HBM residency by keeping the full cache in CPU memory or storage and using $S_l^t$ as both the sparse-attention input and the transfer plan. 
Because $S_l^t$ is available only after query projection, its selection and CPU--GPU transfer must complete before the same layer's attention can begin.

KV prefetch moves block selection ahead of block consumption to hide retrieval latency. 
Layer-ahead prefetching predicts the working set of a later layer and overlaps its transfer with intervening layer computation \cite{infinigen,hisparse}, whereas step-ahead prefetching prepares $\hat{S}_{l}^{t+1}$ during step $t$ and overlaps the transfer with the current decode step \cite{specache,freekv}. 
When the prefetching time is less than the model forward pass time, we can hide data transfer overhead behind the critical path.
At the same time, KV prefetch also requires accurate prediction on cross-layer/step attention patterns, and scalable full-KV offloading that can work with different off-GPU interconnect bandwidth.

\subsection{C1: Sparse Attention Demands Accurate Prefetching}
\label{sec:mot-prefetch}

The first challenge is prediction accuracy.
Prefetching hides latency only when the predicted blocks match those that the next step actually attends to.
A prediction error forces the system either to omit a needed block, reducing accuracy, or to fetch it on demand on the critical path.
Existing predictors fall at two extremes.
Model-integrated designs train dedicated sparse-attention or lookahead modules end-to-end \cite{sparda,dsv4LSA,ji2026echokvefficientkvcache}, which predict well but incur training costs and bind deployment to a dedicated model.
Training-free designs instead reuse an inference-time signal, most commonly the current query as a proxy for the next, trading accuracy for generality \cite{freekv,specache,clo2025}.
We observed that this previous-token proxy recovers the true top-20 blocks with only 84\% average accuracy and substantial layer-to-layer variation on Qwen3-8B (\Cref{sec:appendix-previous-vs-draft-token}).

A general, high-throughput serving system therefore requires a prediction signal that is accurate, training-free\footnote{We use \emph{training-free} from the deployer's perspective: deploying \sys{} requires no predictor-specific architecture design, training, or tuning.
Although the draft model or MTP module may itself be trained, such components are increasingly released as reusable artifacts \cite{eagle3,angelspec2026}.}, and cheap enough to derive and score one full step ahead of decoding.
Such a signal lets sparse prefetching exploit the full-step window; \Cref{sec:design-lookahead} describes how \sys produces it. %
Prediction determines which blocks to fetch, but it does not eliminate the complete KV cache that must remain reachable or the traffic of moving selected blocks across off-GPU interconnects.

\subsection{C2: Host-Memory Capacity Bounds Decode Batch Size}
\label{sec:background-host-memory}
\label{sec:mot-capacity}

The second challenge is the capacity required to retain the entire KV cache outside HBM.
Sparse retrieval and prefetching reduce the HBM usage each request needs while retaining its full KV cache in host memory \cite{freekv,specache,dualpath,hisparse}.
This design lets HBM admit a larger live batch, but host DRAM must back the full context of every request in that batch.
Thus, sparsity reduces the KV accessed at each step, but total KV capacity of all requests can exceed what host DRAM provides. 
Additional DRAM can be provided through larger local configurations or CXL-based memory expansion \cite{nvidia-vera-memory,beluga}, but these approaches increase expensive memory costs and require changing server hardware.

\begin{figure}[t]
	\centering
	\includegraphics[width=\columnwidth]{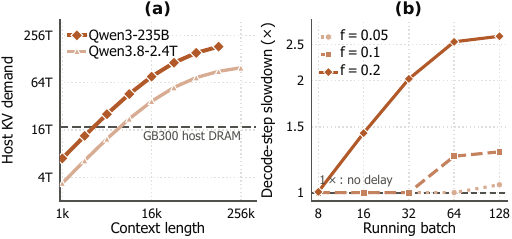}
	\caption{Capacity and bandwidth limits of sparse KV offloading.
	(a): Host KV demand of the maximum HBM-supported sparse batch on GB300 NVL72; the dashed line marks its 17.3\,TB host DRAM.
	(b): Measured decode-step slowdown under overlapped layer-granular PCIe prefetch for Qwen3-8B with a 2{,}048-token working set; $f$ is the nonresident fraction and $1\times$ denotes TPOT when $f=0$.}
	\label{fig:mot-walls}
\end{figure}

For in-decode KV offloading, HBM holds a working set of $W$ tokens and a KV-cache metadata per live request, while the host retains full $L$-tokens KV cache.
Here we use \sys's setup to illustrate the demand of off-GPU memory capacity.
For block size $B$, the KV cache metadata costs $1/B$ of the full KV because it stores two summary vectors per $B$ keys and keys comprise half of KV (\Cref{sec:design-lookahead}).
The largest live batch and the host DRAM required to back it are
\begin{equation}
\small
	N_{\max} = \frac{M_{\mathrm{usable}}}{s_{\mathrm{KV}}\,(W + L/B)},
	\qquad
	D_{\mathrm{host}} = N_{\max}Ls_{\mathrm{KV}} = \frac{L}{W + L/B}M_{\mathrm{usable}},
	\label{eq:host-demand}
\end{equation}
where $M_{\mathrm{usable}}$ is the HBM remaining after model weights.
The multiply factor on $M_{\mathrm{usable}}$ grows with context and approaches $B$.
At the $B{=}16$ and $W{=}2{,}064$-token working set used in our evaluation (\Cref{sec:eval-setup}), it reaches $12.8$ at a 128K context.
\Cref{fig:mot-walls}(a) calculates this demand on GB300 NVL72.
Its host DRAM capacity is quickly exhausted when the serving context length increases.

In AI serving systems, host DRAM needs to support various usage cases, including prefix caching \cite{mooncake,cachedattention,cachegen,memserve}, KV retrieval \cite{infinigen,freekv,hisparse,kvdrive,spin}, CPU-side model or MoE expert offloading \cite{flexgen,moelightning,moegen}, and checkpoint caching \cite{serverlessllm}.
Furthermore, agentic workloads can add operating-system sandboxes, tool invocation, code execution, and persistent state \cite{osworld,swebench,agent-memory} to this demand.
In PD disaggregation, full KV transfer between prefill and decode nodes further amplifies this demand due to full KV replication.
The prefill or storage side already retains the request's full KV cache for handoff or later prompt cache reuse, while the decode node keeps a second copy throughout the decode phase \cite{splitwise,distserve,dualpath,hisparse}.
A PD disaggregation with scalable KV offloading design must therefore keep every block reachable without requiring a full decode-host replica, motivating \Cref{sec:design-tiers}. %

\subsection{C3: Bandwidth Bounds KV Movement}
\label{sec:mot-bandwidth}

The third challenge is off-GPU interconnect bandwidth.
Even when sufficient capacity is available, every selected nonresident block must cross the CPU--GPU link before attention consumes it.
Under next-step prefetch overlap, transfers from different layers remain streaming in flight over the same interconnect.
If their step-wide transfer time $T_{\mathrm{StepXfer}}$ exceeds contended model-forward time $T_{\mathrm{ModelFwd}}$, the residual $T_{\mathrm{stall}}=\max\{0,T_{\mathrm{StepXfer}}-T_{\mathrm{ModelFwd}}\}$ will stall the decode.

\Cref{fig:mot-walls}(b) reports an emulation of layer-granular KV prefetch that isolates this transfer-induced delay, where $f$ represents the ratio of HBM-resident KV cache updated by the prefetch.
At batch 128, $f{=}0.05$, $0.1$, and $0.2$ raise decode-step wall time by $1.05\times$, $1.29\times$, and $2.63\times$, respectively. %
Overlapping KV prefetch in the next-step time window is practical, but a bandwidth-aware control of aggregated prefetch traffic over CPU-GPU interconnect is necessary to minimize TPOT slowdown.

Memory disaggregation adds a second bandwidth wall beyond decode-node CPU--GPU interconnect.
Decode-local KV retrieval techniques \cite{dualpath,hisparse} assumes that the full KV cache is already available in decode-host memory, so it sparsifies the CPU--GPU KV transfer but not the P--D handoff.
A conventional handoff transfers the complete prompt KV cache before decode begins, making admission traffic grow linearly with context length.
Retaining the full KV cache at the prefill node or a remote memory tier avoids the long admission, but off-GPU KV blocks that become active over the decode phase need to cross the scale-out network.
Therefore, we need to bound admission-time and decode-time KV transfer traffic over the network while delivering each active KV block in time, motivating \Cref{sec:design-network-overhead}. %

%% file: tex/03_design.tex
\begin{figure}[t]
    \centering
    \includegraphics[width=\linewidth]{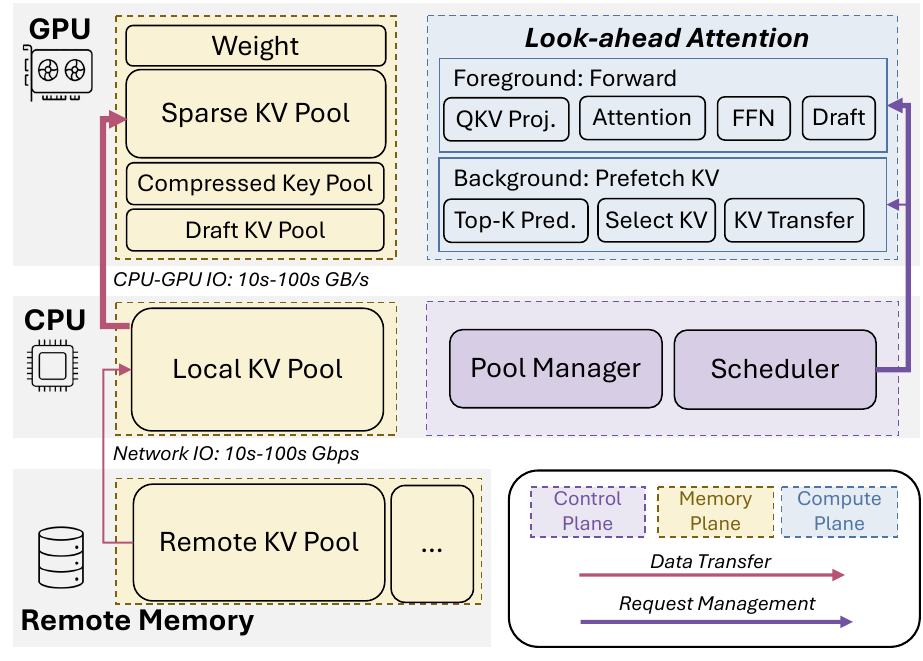}
    \caption{Overview of the \sys{} architecture.
    lookahead attention produces a next-step block ranking, the memory plane retains KV across decode-GPU, decode-CPU, and remote memory, and the control plane manages the KV pools and schedules decoding and prefetching.}
    \label{fig:full-system}
\end{figure}

\section{\sys{} Design}
\label{sec:design}

\sys{} coordinates KV prediction, placement, and movement for long-context decoding (\Cref{fig:full-system}).
Decoding runs over a sparse KV working set while a background pipeline predicts, selects, and prefetches blocks for the next step.
Keeping only this working set, draft KV state, and compressed key summaries in HBM reduces per-request KV storage and allows larger batches.
As batch size grows, storing every request's full KV cache can exhaust decode-host memory (\Cref{sec:mot-capacity}).
\sys{} therefore retains the full prompt KV cache in remote memory and uses decode-host memory as a bounded cache for fetched blocks.
The pool manager maintains the KV pools across devices, and the scheduler coordinates decoding and prefetching at each step.

\subsection{Lookahead Attention}
\label{sec:design-lookahead}

Lookahead attention uses a speculative draft token to construct an approximate next-step query $\hat{q}_l^{t+1}$ during step $t$ (\Cref{sec:background-sparse-prefetch}).
Using this query, \sys{} predicts the per-head block set $\hat{S}_l^{t+1}$, allowing selection and KV transfer for step $t{+}1$ to overlap with the current decode step.
\sys{} remains compatible with speculative decoding, as evaluated in \Cref{sec:reasoning-workload}.

In practice, the GPU holds only each layer's resident working set, not the full KV history, and two problems follow from this.
The first problem concerns the draft query $\hat{q}_l^{t+1}$.
The draft model supplies only a token; to turn it into per-layer queries, the target model must propagate it over the resident working set.
That working set was selected for the normal token, not the draft token, so the draft query it produces carries prediction error.
The second problem concerns top-$K$ selection itself.
Even with an accurate draft query, the resident set carries no information about nonresident blocks, so it cannot rank the full context.
\sys{} resolves the first with an empirical observation and the second with a small amount of metadata in HBM.

\begin{figure}[t]
    \centering
    \includegraphics[width=\linewidth]{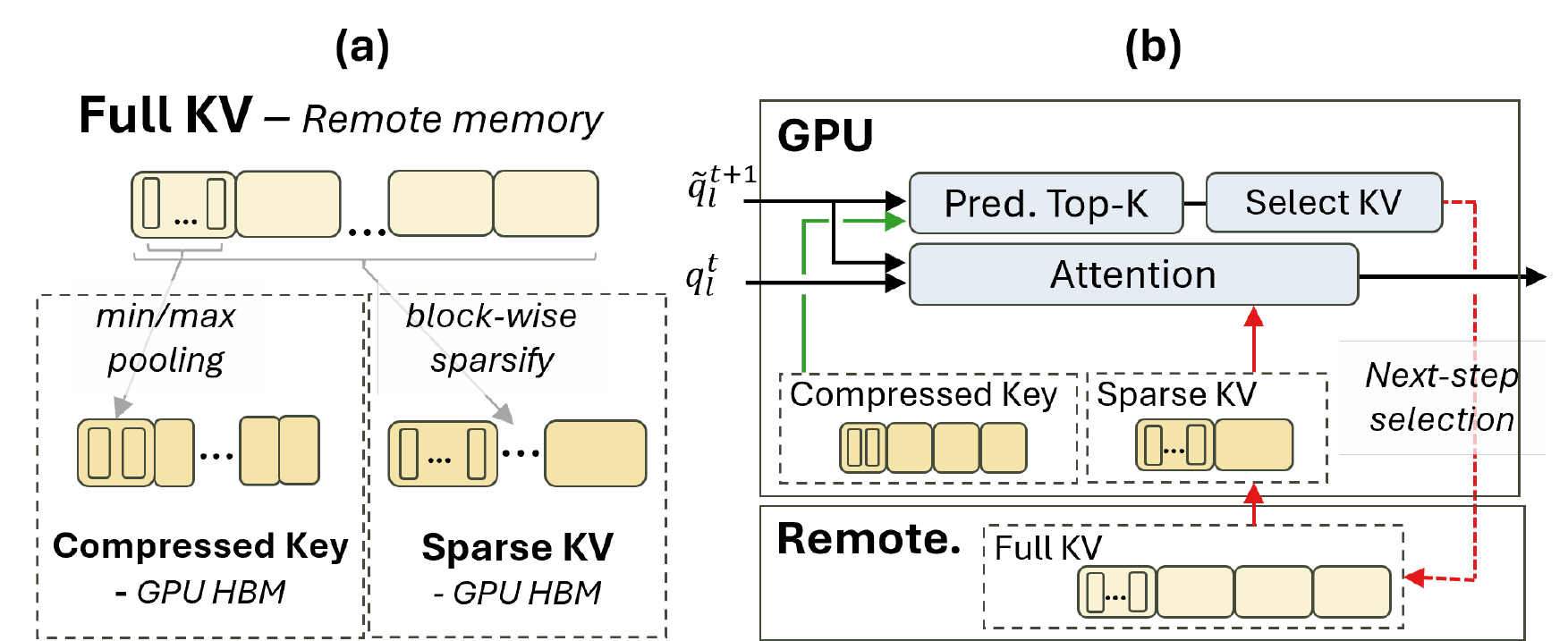}
    \caption{lookahead attention.
    (a) The GPU stores per-block min/max key summaries and a sparse subset of the full KV cache.
    (b) The attention kernel processes the normal and draft queries together over this subset; the draft query ranks the summaries to predict the next per-head top-$K$ set.}
    \label{fig:lookahead-attn}
\end{figure}
Prior work \cite{freekv} observes strong temporal locality in block importance across adjacent decode steps, so the normal token's resident blocks are likely to contain the next step's top-$K$ set.
\sys{} therefore propagates the draft token directly over the normal token's sparse KV.
Our profiling of Qwen3-8B on GSM8K with Eagle-3 \cite{eagle3} finds that the resulting top-$K$ set agrees with the exact next-token set by $98.28$--$99.69\%$ across all 36 layers, averaging $98.74\%$ (\Cref{sec:appendix-draft-token-prediction}).

To rank nonresident blocks, \sys{} keeps the selection metadata required by the sparse-attention algorithm in HBM.
As a concrete case, our prototype adopts Quest-style summaries, which store the coordinate-wise minimum and maximum keys, $k_b^{\min}$ and $k_b^{\max}$, for each logical block \cite{quest} (\Cref{fig:lookahead-attn}).
Each draft query scans these summaries and ranks all logical blocks independently for every KV head,
so predictions cover the full context without restoring any keys.
Because a summary holds two vectors per block instead of $B$ keys, the summaries require only a small fraction of the full KV cache's capacity and per-step read traffic.
We express these summaries as a center $c_b = (k_b^{\min}+k_b^{\max})/2$ and a radius $r_b = (k_b^{\max}-k_b^{\min})/2$.
The score of a block is the sum over the query heads $h$ that share this KV head:
\begin{equation}
    s_b(\alpha) = \sum_h \bigl( q_h^{\mathsf T} c_b + \alpha\, |q_h|^{\mathsf T} r_b \bigr),
    \label{eq:score}
\end{equation}
$\alpha{=}{-1}$ gives a lower bound on the summed query--key scores within a block, while $\alpha{=}1$ gives an upper bound.
We use $\alpha{=}0.5$ based on the ablation in \Cref{sec:appendix-alpha}.

Also shown in \Cref{fig:lookahead-attn}, each layer now runs one path for both tokens.
The QKV projection produces the normal and draft queries together.
The attention kernel takes the stacked queries $\{q_l^{t}; \hat{q}_l^{t+1}\}$ over one resident KV working set and returns both outputs in one call.
The draft query then drives top-$K$ prediction, which scans the compressed summaries and ranks the blocks the next step will need.
Selection turns that ranking into the blocks to fetch.
Both mechanisms add little to the foreground: shared execution reads the model weights and the resident KV once for both tokens, and the summary scan reads only two vectors per block.

\subsection{Tiered KV Placement}
\label{sec:design-pool}
\label{sec:design-tiers}

\begin{figure}[t]
    \centering
    \includegraphics[width=\linewidth]{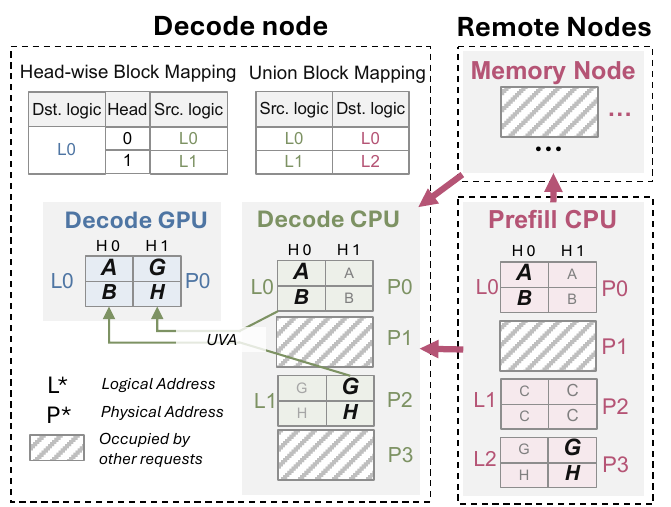}
    \caption{Tiered KV pool and hierarchical block mappings.
    Remote memory (pink) retains the full prompt KV, decode-host memory (green) caches whole blocks selected across heads, and HBM (blue) holds a fixed per-head KV working set.
    Letters identify tokens, and columns $H_0$ and $H_1$ show their KV entries for the two heads.
    Bold entries are selected by the corresponding head for attention, while gray entries are stored but unselected.}
    \label{fig:kv-tiers}
\end{figure}

Disaggregated serving separates prefill and decode workers, while KV-storage systems store KV caches in dedicated memory tiers \cite{splitwise,distserve,mooncake,dualpath,memserve,beluga}. 
The complete prompt KV cache can therefore remain in prefill-host memory or remote memory.
\sys{} fetches selected KV blocks from these existing caches, so the decode node does not need to store a full copy of each request's KV cache in host memory.
\sys{} organizes (i) decode-GPU HBM, (ii) decode-host memory, and (iii) remote memory into a tiered KV pool for sparse attention (\Cref{fig:kv-tiers}).

\subsubsection{Per-Tier Residency}
\label{sec:design-decode-residency}
\label{sec:design-custody}

\emph{(i) Decode-GPU HBM} maintains exactly $K$ blocks per KV head. 
Newly generated blocks replace existing entries independently for each KV head.
Each head maintains its own working set, so different heads can hold different token sets while GPU residency remains fixed.
This flexibility comes at the cost of finer-grained KV transfers from host memory.
To minimize this overhead, our implementation uses a unified virtual addressing (UVA~\cite{cuda-memory}) gather kernel running on GPU SMs.
The kernel reads the selected per-head KV entries directly from pinned host memory over PCIe and writes them into HBM.

\emph{(ii) Decode-host memory.} \sys{} fetches missing blocks from the union of all heads' top-$K$ selections into the host cache.
Even if only one KV head selects a missing block, \sys{} fetches the entire block from remote memory.
Remote fetching involves CPU gather, network transfer, and scatter; gathering individual head slices can become latency-bound due to small, non-contiguous reads.
Fetching whole blocks amortizes CPU gather overhead over larger reads and enables reuse across heads and subsequent decoding steps.
Generated KV remains on the decode node and is retained in decode-host memory for the request's lifetime, setting a minimum per-request host-memory footprint.

\emph{(iii) Remote memory} retains the complete prompt KV cache in registered memory outside the decode host.
In PD disaggregation, the complete prompt KV cache initially resides in prefill-host memory; when its capacity is exhausted, \sys{} offloads complete prompt KV caches to additional registered pools, including dedicated memory servers.
Retaining complete prompt KV blocks in remote memory preserves prefix reuse across subsequent requests.

In single-node serving, local host memory retains the complete KV history and supplies selected blocks directly to the GPU.

\subsubsection{Hierarchical Block Mapping}
\label{sec:design-headwise}
\label{sec:design-serving}

\sys{} builds on the paged KV cache of vLLM, which stores each request's KV in fixed-size pages, every page holding the same token range for all KV heads \cite{vllm-pagedattention}.
As new tokens arrive, new pages are appended and existing pages are never rewritten.
Sparse decode breaks both assumptions: (i) different heads have different top-$K$ patterns and thus may need different token ranges in one page; (ii) a fixed token budget cannot grow but must still accept new tokens as they come in.

\sys{} keeps every tier's logical-to-physical memory mapping unchanged and adds two logical-to-logical tables above them (\Cref{fig:kv-tiers}), which illustrates the mappings with two KV heads: $H_0$ selects the block containing tokens $(A,B)$, while $H_1$ selects $(G,H)$.

\emph{Union block mapping} records where each selected remote block is cached on the decode host.
The prefill host stores $(A,B)$ at logical block $L_0$ and $(G,H)$ at $L_2$; the decode host caches them as $L_0$ and $L_1$, respectively.
Both host blocks retain KV entries for all heads, while the unselected $(C,D)$ block remains remote.
\emph{Head-wise block mapping} allows one GPU logical block to hold different source blocks for different heads.
For GPU block $L_0$, the table selects decode-host block $L_0$ for $H_0$ and decode-host block $L_1$ for $H_1$.
UVA gather copies the corresponding head slices into GPU physical page $P_0$, placing $(A,B)$ under $H_0$ and $(G,H)$ under $H_1$.
The existing page tables resolve logical indices to physical pages independently within each tier.

Changes in top-$K$ selections or block residency require only lightweight updates to the corresponding mapping entries, without moving unrelated page contents or rewriting the underlying page tables.
Because GPU pages retain their original layout, sparse requests use the same FlashAttention kernel and batched attention call as dense requests.
Under tensor parallelism, each GPU updates only the head-wise entries for its local heads and requires no cross-GPU synchronization.

\subsubsection{Relieving Decode-Host Memory Pressure}
\label{sec:design-host-capacity}

As discussed in \Cref{sec:mot-capacity}, decode-host memory has finite capacity and cannot retain all KV blocks as context lengths and batch sizes grow.
\emph{Decode-host LRU} bounds decode-host residency while retaining whole blocks shared across heads.
Without eviction, remote fetches could eventually accumulate the full prompt KV cache in decode-host memory.
\sys{} therefore uses a per-request budget of $f_{\mathrm{Host}}P+F$ blocks, where $P$ is its prompt length in blocks, $f_{\mathrm{Host}}$ is the \emph{decode-host residency ratio} for prompt KV, and $F$ reserves space for generated KV.
For this budget to bound residency, $F$ must cover the request's maximum number of generated blocks, and the total budget must accommodate all generated blocks and the protected current working set.
A periodic pass evicts the request's least recently used prompt blocks while protecting generated blocks and the current working set.
The evicted prompt blocks remain available in remote memory, so this policy does not approximate attention; a later selection fetches them again.
Smaller $f_{\mathrm{Host}}$ therefore increases concurrency at the cost of additional network traffic and possible fetch stalls, as evaluated in \Cref{sec:eval-ablation}.

\subsection{Asynchronous Prefetching}
\label{sec:design-memory}
\label{sec:design-network-overhead}

lookahead attention at step $t$ predicts each layer's block demand for step $t{+}1$.
\sys{} prefetches the selected blocks from remote memory through decode-host memory to HBM using a four-stage pipeline of prediction, selection, remote fetch, and local fetch.
\sys{} overlaps these stages across layers to hide prefetch latency and uses HBM-side LRU to bound per-step transfer traffic.

\subsubsection{Per-Layer Prefetch Pipeline}
\label{sec:design-transfer-overhead}

\begin{figure}[t]
    \centering
    \includegraphics[width=\linewidth]{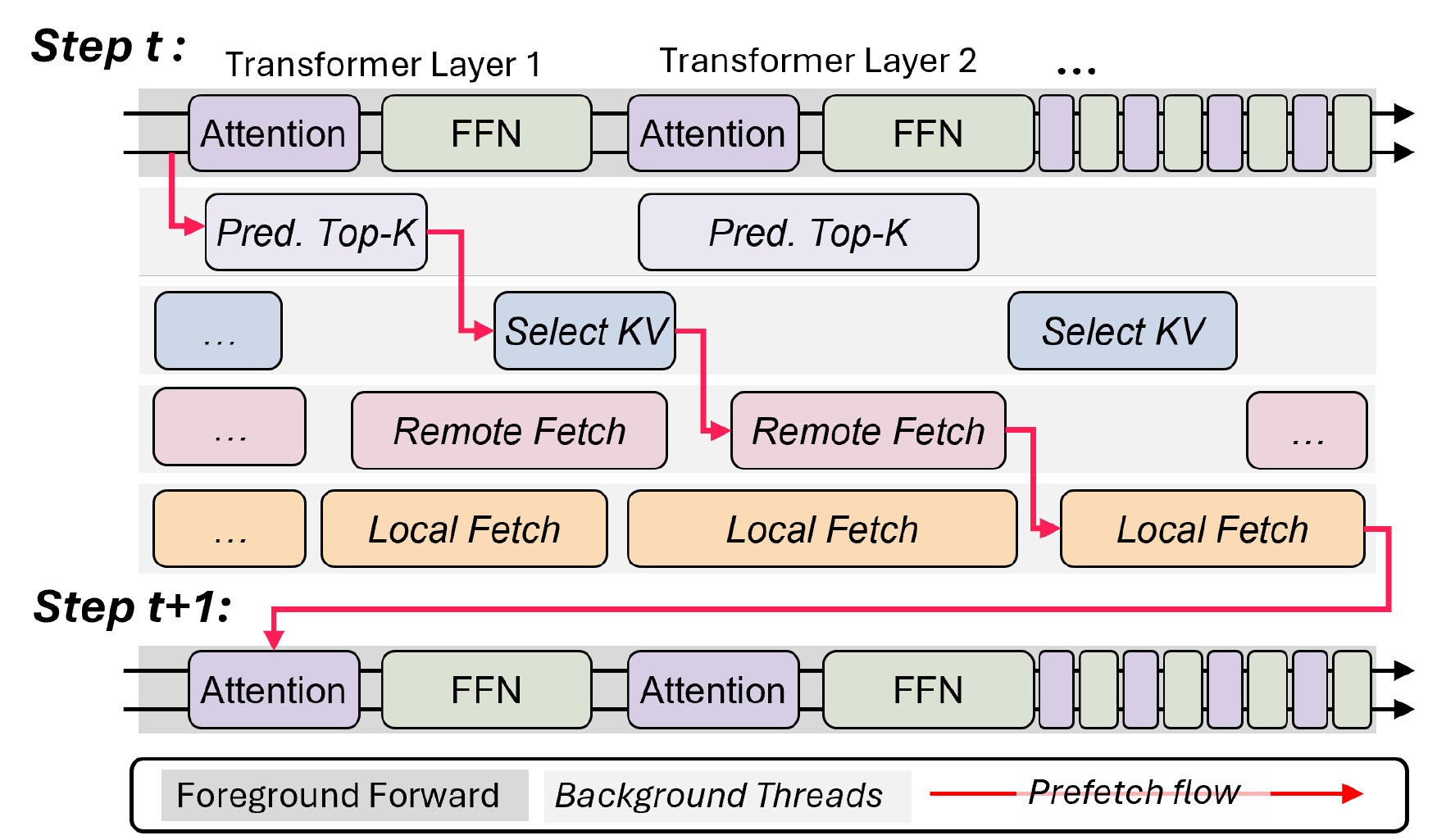}
    \caption{The asynchronous per-layer prefetch pipeline across two decoding steps.
    Red arrows follow one layer's prefetch task from prediction at step $t$ until its KV is ready for attention at step $t{+}1$.
    Different layers overlap across the four stages.}
    \label{fig:prefetch-pipeline}
\end{figure}

\Cref{fig:prefetch-pipeline} illustrates the per-layer prefetch pipeline.
For each layer, \sys{} constructs one prefetch task from the draft query at step $t$ to prepare the KV blocks needed at step $t{+}1$.
The prediction stage (\ie{} Pred. Top-K in \Cref{fig:prefetch-pipeline}) ranks compressed keys and produces the predicted per-head block set $\hat{S}_l^{t+1}$.
The selection stage (\ie{} Select KV in \Cref{fig:prefetch-pipeline}) applies the transfer budget and classifies the selected blocks by residency: HBM hits require no transfer, decode-host hits require only local fetch, and decode-host misses require remote fetch followed by local fetch.
Remote fetch coalesces whole-block misses across requests for layer $l$ and caches the fetched blocks in decode-host memory; local fetch then uses a single UVA gather to copy the selected per-head slices from pinned host memory into HBM.

\sys{} runs the stages in background workers, with CUDA events preserving dependencies within each layer while allowing different layers to occupy different stages concurrently.
Before layer $l$'s attention at step $t{+}1$, the foreground waits only for that layer's local fetch to complete, allowing other layers' prefetch tasks to proceed independently.

In PD-disaggregated serving, \sys{} initializes each request by transferring its initial selected KV blocks, compressed-key cache, and draft KV state from the prefill node to the decode node before decoding begins.
In non-PD serving, prefill and decode share a node whose host memory already retains the complete KV cache, so remote fetch is bypassed.

\subsubsection{Relieving Off-GPU Bandwidth Pressure}
\label{sec:design-bandwidth-overhead}

Fine-grained asynchronous pipelining hides fetch latency and exposes the bandwidth cost of working-set updates.
Rapid changes in the top-$K$ sets can still exceed the per-step PCIe budget.

\emph{HBM-side LRU} bounds PCIe traffic while retaining exactly $K$ blocks per head.
The \emph{HBM fetch ratio} $f_{\mathrm{HBM}}$ specifies the maximum fraction of each head's KV working set that can be replaced per decode step.
At each layer, selection admits at most $f_{\mathrm{HBM}}K$(\ie, HBM fetch-cap) predicted nonresident blocks per head and evicts the same number of least recently used resident blocks.
Across $H$ KV heads, this limits local transfer to $f_{\mathrm{HBM}}KH$ block-head entries per layer.
We choose $f_{\mathrm{HBM}}$ so that the total PCIe transfer time across all layers and requests fits within one model-forward pass (\Cref{sec:mot-bandwidth}).
The admission cap applies after the first working set has been fully seeded.
However, when more than $f_{\mathrm{HBM}}K$ blocks change, some predicted top-$K$ blocks do not enter HBM and older blocks remain in their place.
This deviation can omit important attention blocks and reduce accuracy, which \Cref{sec:eval-ablation} evaluates.

%% file: tex/04_evaluation.tex
\begin{figure*}[t!]
\centering
\includegraphics[width=\linewidth]{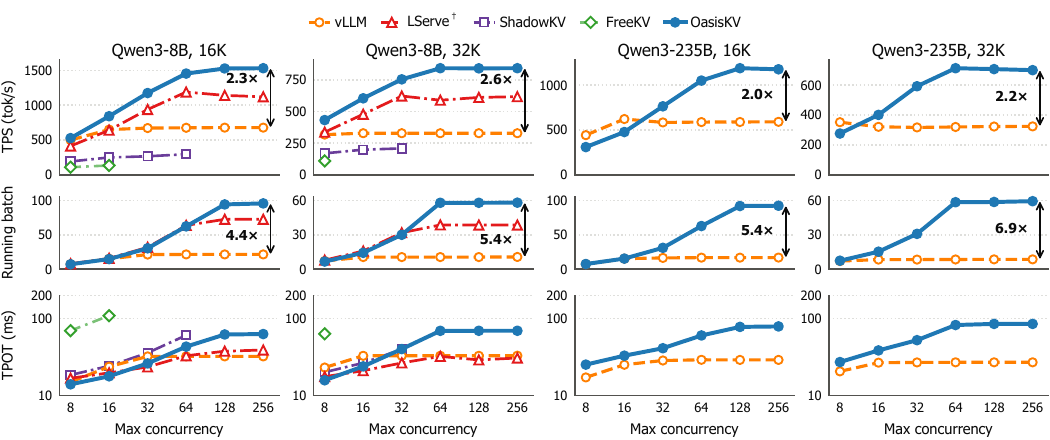}
\caption{Synthetic decode sweep over max concurrency at 16K and 32K context, on
Qwen3-8B (TP1) and Qwen3-235B (TP8). We compare \sys{} against dense vLLM, LServe\textsuperscript{\textdagger} \cite{yang2025lserve}, ShadowKV ~\cite{shadowkv}, and FreeKV~\cite{freekv}.
LServe\textsuperscript{\textdagger} combines sparsity and 8-bit KV quantization, weights with FP16; all other sparse baselines use sparsity only.
Rows report TPS, running batch, and TPOT; each request generates 2{,}048 tokens.
 Arrows mark the \sys{}/vLLM ratio at max concurrency 256.}
\label{fig:synth}
\end{figure*}

\section{Evaluation}
\label{sec:eval}

\subsection{Experimental Setup}
\label{sec:eval-setup}
\paragraph{Testbed.}
All experiments run on two nodes with eight NVIDIA H100 GPUs, each with 80\,GB of HBM3, two Intel Xeon Platinum 8480C CPUs with 112 cores across two NUMA nodes, and 2\,TB of host memory.
The GPUs are connected by NVLink and NVSwitch, and each GPU connects to its local CPU socket over PCIe Gen5 $\times$16.
Single-GPU experiments use one H100, while multi-GPU experiments use all eight under tensor parallelism.
The PD-disaggregated experiments use two such nodes, connected by NVIDIA ConnectX-7 NICs with 400-Gbps Ethernet ports and RoCE support.

\paragraph{Models.}
We evaluate Qwen3-8B, Qwen3-235B-A22B (MoE), and Llama-3.1-8B-Instruct, covering dense and MoE models from 8B to 235B parameters.
We use publicly available EAGLE-3 draft heads from Hugging Face for each target model~\cite{eagle3-qwen3-8b-hf,eagle3-qwen3-235b-hf,eagle3-llama31-8b-hf}.
Qwen3 thinking-mode runs use the recommended temperature of 0.6, top-$p$ of 0.95, and top-$k$ of 20.
LLama-3.1-8B-Instruct uses greedy decoding.

\paragraph{Workloads.}
For accuracy, we use AIME24, AIME25, GPQA-Diamond for long-output benchmark, and LongBench~v2 for 0-shot long-input benchmark.
The three reasoning benchmarks use Qwen3 thinking mode, with up to 38{,}912 generated tokens for AIME and 32{,}768 for GPQA-Diamond.
When an input exceeds the model's context window, we apply the chat template, retain the first and last halves of the allowed tokens, and remove the middle.
LongBench~v2 groups its questions by context length, so we evaluate the short (180), medium (215), and long (108) splits separately and report the sample-weighted average as the overall score.
For performance, synthetic workloads sweep input/output length, batch size, and request rate, while the AIME runs measure end-to-end performance on the same requests used for accuracy.
For the 32K PD workload, we construct 128 requests in 32 four-request groups, each targeting a 32K-token prompt and sharing a prefix of $h\times32$K tokens. 
Before measurement, one request per group populates the cache. 

\paragraph{Implementation}
On top of the V1 engine of vLLM v0.12.0,
\sys{} extends the engine with a sparse-attention backend, a GPU model runner, a KV cache manager, and scheduler support for the transition from dense to sparse attention.
We implement compressed-key updates, head-wise top-$K$ prediction, hierarchical block mapping, and KV transfer in C++ and CUDA.
Persistent C++ workers dispatch Top-$K$ Prediction, KV Selection, and KV Transfer on separate background CUDA streams.
Model execution uses the foreground stream, and CUDA events enforce the dependencies among the four streams for each layer.
The full KV cache resides in pinned CPU memory.
A custom unified virtual addressing (UVA)~\cite{cuda-memory} gather kernel reads the selected block-head entries from this cache over PCIe and writes them directly into the resident GPU pages.
For PD disaggregation, we use NIXL 1.3.0 \cite{NIXL} over UCX 1.21.0 \cite{ucx} for KV cache transfer between nodes.
\begin{figure*}[t]
\centering
\includegraphics[width=\linewidth]{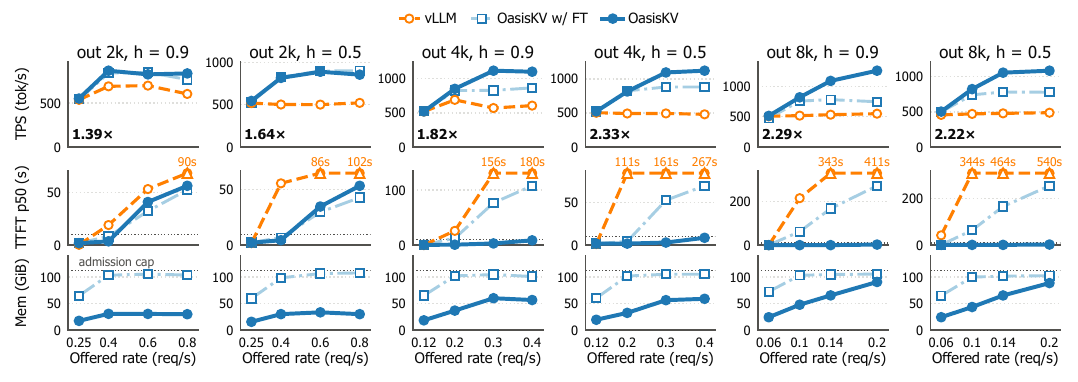}
\caption{PD-disaggregated Qwen3-8B serving with 32K-token prompts under 125\,GiB host budgets.
Columns vary output length and shared-prefix fraction $h$; rows report throughput, median TTFT, and avg decode-host memory usage.
vLLM and \sys w/ FT denote full transfer for dense and sparse. Label report \sys{}/vLLM at each column's last offered rate.}
\label{fig:pdserving}
\end{figure*}

\paragraph{Baselines and configurations.}
For end-to-end performance, we compare \sys{} with dense vLLM using FlashAttention3 and a GPU-resident full KV cache.
Dense vLLM uses full CUDA graphs, while \sys{} uses piecewise CUDA graphs.
For single-GPU system performance, we compare against LServe~\cite{yang2025lserve}, ShadowKV~\cite{shadowkv}, and FreeKV~\cite{freekv}, each on its own serving stack.
By default, we use BF16 for both weights and KV cache.
LServe is a sparse-attention baseline, but also applies 8-bit KV cache; we mark it with a dagger in \Cref{fig:synth}.
FreeKV performs KV prefetch, while ShadowKV performs KV retrieval.
For accuracy, we compare \sys{} with full attention, Quest~\cite{quest} representing sparse attention, and FreeKV~\cite{freekv} representing existing KV prefetch.
We use the implementations of both methods from FreeKV's repository.
Quest and FreeKV attend to 2{,}048 tokens in each sparse layer: 128 sink tokens, 128 recent tokens, and 1{,}792 dynamically selected tokens.
They keep the first layer dense.
\sys{} uses a block size of $B{=}16$ and selects $K{=}128$ blocks (\ie 2{,}048 tokens) per KV head, without any manually selected sink tokens or dense layers.
Each head also retains the active local block, so \sys{} attends to 129 blocks, or 2{,}064 tokens.
The compressed-key cache stores two summary rows per block and occupies $1/16$ of the full KV-cache size.
We use $f_{HBM}=0.05$ and $f_{Host}=0.2$ for eviction strategy.

\subsection{System Performance}
\label{sec:eval-sysperf}

\subsubsection{Single-Node serving}
\label{sec:single-node}
\label{sec:eval-synthetic}
\Cref{fig:synth} sweeps max concurrency at 16K and 32K with 2{,}048-token outputs.
On Qwen3-8B, \sys{} reaches 1{,}532 vs.\ 677 tok/s at 16K and 842 vs.\ 328 at 32K at max concurrency 256, gains of $2.3\times$ and $2.6\times$ over vLLM.
At concurrency 16 and 16K, it also improves TPS from 649 to 841 while reducing TPOT from 23.6 to 17.7\,ms.
At high concurrency, its bounded KV working set supports 96.1 running requests at 16K, compared with 21.7 for vLLM, sustaining TPS even when its TPOT becomes higher.
\sys{} also exceeds LServe\textsuperscript{\textdagger} by $1.21$--$1.43\times$ in TPS and supports $1.32$--$1.51\times$ its running batch at max concurrency.
Other sparse attention systems fall further behind; missing points denote OOM or unsupported configurations.
Qwen3-235B shows a similar latency throughput trade-off under TP8.
Its KV is a small share of GPU memory, so sparsity saves less while our extra per-step overhead remains --- at low batch our TPOT is therefore higher than dense. 
Even so, we overtake dense at $B{=}32$ (16K) and $B{=}16$ (32K), reaching up to $1.9\times$ TPS. 

Overall, this result shows that \sys{} uses a fixed HBM budget more efficiently, achieving lower TPOT before saturation and supporting a larger running batch once dense serving reaches its capacity limit.
Qwen3-8B benefits from both effects, whereas Qwen3-235B gains primarily from the larger batch size after dense KV residency becomes HBM-capacity limited.
The consistent improvement over baselines further shows that lookahead attention and asynchronous KV prefetch allow \sys{} to scale with concurrency while introducing minimal overhead.

\subsubsection{Multi-node serving} 
\label{sec:eval-pd}

\Cref{fig:pdserving} reports the PD-serving results.
The Qwen3-8B workload uses 32K-token prompts, 2K, 4K, and 8K outputs, and shared-prefix fractions $h\in\{0.5,0.9\}$.
Prefill and decode each run on one H100, with a 125\,GiB host KV budget and 480\,GiB remote memory. 
We emulate the remote tier using another NUMA node DRAM on the prefill server, whose measured cross-socket bandwidth matches a 400-Gbps RoCE link; decode-to-tier reads still traverse the physical RoCE fabric.
We use a  small host memory size relative to system configuration for demonstrating that \sys can break the capacity limitation at decode-host memory.
We compare vLLM, \sys{} w/ Full Transfer (FT), and \sys{} under the same offered-rate sweep.
Dense vLLM uses automatic prefix caching, while \sys{} w/ FT and \sys{} use the same \sys{}'s tiered KV management.

At each panel's last offered rate, \sys{} achieves $1.39$--$1.64\times$, $1.82$--$2.33\times$, and $2.22$--$2.29\times$ vLLM TPS for 2K, 4K, and 8K outputs, respectively.
At 8k and 0.2\,req/s, \sys{} reaches 1{,}257 and 1{,}085 tok/s for $h=0.9$ and $h=0.5$ compared to 549.6 and 488.1 tok/s for vLLM.
Also, it exceeds \sys{} w/ FT by $1.68\times$ and $1.39\times$.
At this point, it sustains 46.7--47.1 active requests, compared with 14.8--19.4 for vLLM and 20.9--21.0 for \sys{} w/ FT.
\sys{} uses 88.9--91.0\,GiB of decode-host memory, compared with 102.8--105.8\,GiB for \sys{} w/ FT.
The resulting admission headroom reduces median TTFT to 3.36\,s against 270\,s for \sys{} w/ FT and 540\,s for vLLM.
\sys{}'s advantage over other baselines grows with output length because longer generations keep more requests active and expose the decode node's capacity limit.
These results show that \sys{} converts the same host capacity into higher concurrency along with minimizing TTFT and decode-host mem usage.

\begin{figure}[t]
\centering
\includegraphics[width=\linewidth]{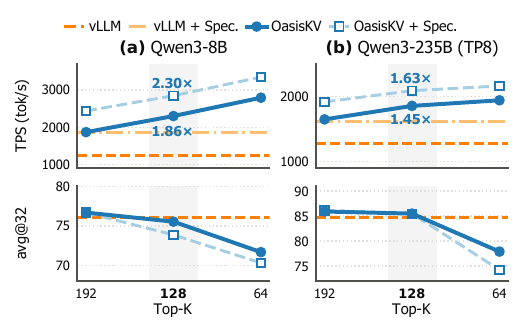}
\caption{Top-$K$ and speculative decoding sweep study on Qwen3-8B and Qwen3-235B-A22B at AIME24. Labels report its TPS speedup at $K=128$ over vLLM.
}
\label{fig:sysperf}
\end{figure}

\input{figures/04_accuracy.tex}

\subsubsection{Reasoning workload}
\label{sec:reasoning-workload}

\Cref{fig:sysperf} evaluates end-to-end AIME24 serving across the Top-$K$ budget and speculative-decoding configuration to see the trade-off between accuracy and TPS.
Reducing $K$ shrinks the resident working set and increases TPS, but eventually loses accuracy.
With \sys{}, reducing $K$ from 192 to 64 raises the speedup over dense vLLM from $1.51\times$ to $2.26\times$ on Qwen3-8B and from $1.29\times$ to $1.51\times$ on Qwen3-235B.
Over the same sweep, avg@32 falls from 76.67 to 71.67 and from 85.94 to 77.92, respectively.
At the default $K{=}128$, \sys{} reaches 2{,}300 vs.\ 1{,}235 tok/s on Qwen3-8B and 1{,}856 vs.\ 1{,}283 tok/s on Qwen3-235B, while remaining close to dense accuracy. %
It also outperforms vLLM + Spec. drafting same single token for acceptance.
These results support $K{=}128$ for the default throughput without sacrificing accuracy.

In \sys{} + Spec., \sys{} takes two draft tokens. The first is for speculative acceptance, and the second is always rejected.
\sys{} unifies the two drafts' top-$K$: it averages the two draft queries and scores the compressed keys with the averaged query.
At $K{=}128$, this composition further achieves 2{,}846 tok/s on Qwen3-8B and 2{,}088 tok/s on Qwen3-235B, corresponding to $2.30\times$ and $1.63\times$ speedups over dense vLLM.
It leaves Qwen3-235B accuracy unchanged and slightly lowers Qwen3-8B accuracy to 73.85.
The two techniques can therefore compose orthogonally: speculative decoding retains its optimization purpose while supplying \sys{} with a lookahead prefetch signal.

\subsection{Accuracy Compared with Prior Work}
\label{sec:eval-accuracy}

\sys{} introduces both prefetching and eviction policies at HBM that may affect model accuracy.
Therefore, we compare \sys{} with Quest~\cite{quest} and FreeKV~\cite{freekv} under the same 2{,}048-token KV budget.
Because the baselines run in HuggingFace Transformers and \sys{} runs in vLLM, \Cref{tab:accuracy} compares each method only with full attention in the same stack.
Following FreeKV's protocol, long-input runs use one seed, while long-output runs use eight samples per AIME question and four per GPQA-Diamond question.
For long-input benchmark, \sys{} remains within 0.61 points loss against full attention across two models on LongBench~v2, which performs similarly as Quest and FreeKV.
For long-output benchmark, we evaluate only Qwen3-8B because Llama-3.1-8B-Instruct is not a reasoning model.
Across three reasoning workloads, \sys{}'s overall scores are noticeably better than two other methods.
Per-benchmark results shows varied small impact on avg@$k$ and pass@$k$.
Despite this variation, \sys{} has the smallest aggregate deviation for both metrics.

\begin{table}[t]
\caption{HBM fetch-cap ($f_{\mathrm{HBM}}$) ablation on Qwen3-8B (AIME24, LRU eviction). The cap bounds the
blocks fetched per step by a fetch ratio. \textbf{Bold} marks our default
operating point. full attention scores 76.04 / 90.00. $\uparrow$ and $\downarrow$ indicate higher and lower is better, respectively.}
\label{tab:eviction}
\centering
\small
\setlength{\tabcolsep}{5pt}
\resizebox{\columnwidth}{!}{%
\begin{tabular}{@{}lrrrcc@{}}
\toprule
$f_{\mathrm{HBM}}$ & Fetch & BW & TPS & \multicolumn{2}{c}{AIME24} \\
\cmidrule(l){5-6}
          & (GB/step; $\downarrow$) & (GB/s; $\downarrow$) & (tok/s; $\uparrow$) & avg@32 ($\uparrow$) & pass@32 ($\uparrow$) \\
\midrule
0.02          & 0.60          & 11.3           & 2{,}386          & 75.00          & 86.67 \\
\textbf{0.05} & \textbf{1.49} & \textbf{26.3} & \textbf{2{,}300} & \textbf{75.52} & \textbf{93.33} \\
0.10          & 2.87          & 31.4          & 1{,}421          & 76.77          & 86.67 \\
0.20          & 4.34          & 34.0          & 1{,}035          & 77.40          & 93.33 \\
\midrule
Fetch all     & 5.05          & 33.5          & 824              & 76.46          & 86.67 \\
\bottomrule
\end{tabular}
}
\end{table}

\subsection{Analysis}
\label{sec:eval-ablation}
\subsubsection{Eviction Strategy}

\sys{} controls host-memory capacity and off-GPU bandwidth by two LRU eviction strategies.
\Cref{tab:eviction} ablates $f_{\mathrm{HBM}}$ on Qwen3-8B.
Across the sweep, relaxing the cap raises per-step traffic from 0.60 to 5.05\,GB and reduces throughput from 2{,}386 to 824\,tok/s, while avg@32 remains within 75.00--77.40.
Because attention computation is unchanged, this inverse traffic--throughput trend identifies PCIe bandwidth, rather than compute, as the binding resource.
At ratios of 0.10 and above, effective demand saturates at 30--34\,GB/s, yet throughput continues falling from 1{,}421 to 824\,tok/s and accuracy does not improve consistently.
Extra fetching therefore stalls decoding without a commensurate accuracy return.
Our default ratio of 0.05 remains within 0.5 point of dense accuracy while delivering $2.8\times$ the fetch-all throughput.
Similar trade-off is also observed in long-input workloads, shown in \Cref{sec:appendix-eviction-longbench}.

\begin{figure}[t]
\centering
\includegraphics[width=0.95\columnwidth]{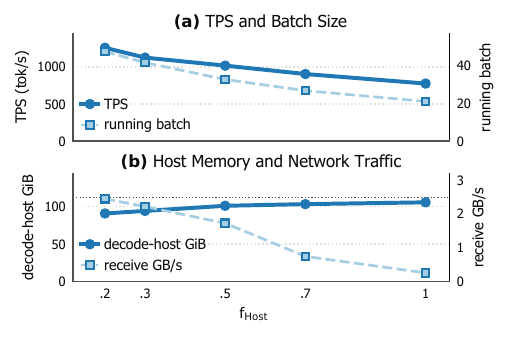}
\caption{Decode-host residency ($f_{\mathrm{Host}}$) ablation on Qwen3-8B serving with 32K-token prompts, 8K outputs and $h{=}0.9$.}
\label{fig:host-residency}
\end{figure}

\Cref{fig:host-residency} ablates $f_{\mathrm{Host}}$, the fraction of each prompt that may remain in the decode-host cache.
At $f_{\mathrm{Host}}{=}0.2$, 91.0\,GiB supports 47.0 active requests, whereas $f_{\mathrm{Host}}{=}1.0$ uses 105.8\,GiB for 20.9 requests.
This $2.25\times$ larger batch raises throughput from 776 to 1{,}257\,tok/s, a $1.62\times$ gain.
The gain costs network bandwidth: evicting more host blocks increases the mean receive rate from 0.26 to 2.45\,GB/s.
Thus, $f_{\mathrm{Host}}$ trades per-request host capacity for decode-time network traffic.

\subsubsection{Network Bandwidth}
\label{sec:eval-partial}

\begin{figure}[t]
\centering
\includegraphics[width=\columnwidth]{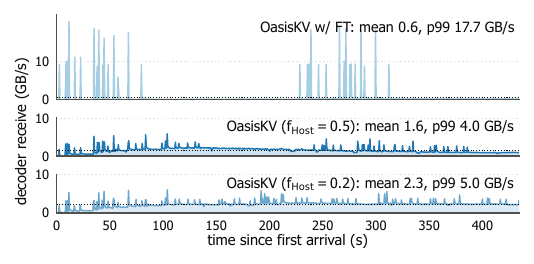}
\caption{Decode-side receive-bandwidth timelines for Qwen3-8B with 32K-token prompts, 8K-token outputs, $h{=}0.9$, 0.2\,req/s, and a 125\,GiB host budget.
Rows compare full transfer with \sys{} at $f_{\mathrm{Host}}{=}0.5$ and $0.2$.}
\label{fig:pdtransfer}
\end{figure}

\Cref{fig:pdtransfer} compares bandwidth timelines at decode worker under the same offered load.
Full transfer leaves the link mostly idle between admission bursts, yielding a 17.7\,GB/s 99th percentile despite a 0.6\,GB/s mean.
By admitting only the predicted union and fetching subsequent misses during decode, \sys{} at $f_{\mathrm{Host}}{=}0.5$ limits the 99th percentile to 4.0\,GB/s while sustaining a 1.6\,GB/s mean.
Reducing $f_{\mathrm{Host}}$ to 0.2 increases remote misses and raises the mean and 99th percentile to 2.3 and 5.0\,GB/s, respectively, but the peak remains $3.5\times$ below full transfer.
Thus, \sys{} converts bursty handoff traffic into a steadier decode-time stream, with host residency controlling its bandwidth.
This is particularly helpful to fit \sys's network traffic into challenging scale-out network environment in AI data centers.
We will extend \sys's potential to prefetch in-decode KV cache over SSD-based storage tier in future work.

%% file: figures/04_accuracy.tex
\providecommand{\dd}[1]{{\color{gray!60!black}#1}}

\definecolor{posgreen}{RGB}{35,120,75}

\providecommand{\ddb}[1]{%
  {\itshape\bfseries\color{posgreen}#1}}

\providecommand{\secrow}[1]{%
  \rowcolor{gray!12}\multicolumn{6}{@{}l}{\itshape #1}}

\providecommand{\dm}[2]{%
  \makebox[4.0em][l]{#1}#2}

\begin{table}[t]
\centering
\caption{Accuracy under the same 2{,}048-token KV budget. Each retrieval method
is read against the full-attention anchor of its own stack: the parenthesized
$\Delta$ is the difference from that stack's Full column (Quest, FreeKV vs.\
Full~(HF); Ours vs.\ Full~(vLLM)), never across stacks.}
\label{tab:accuracy}

\setlength{\tabcolsep}{4pt}
\renewcommand{\arraystretch}{1.1}

\resizebox{\columnwidth}{!}{%
\begin{tabular}{@{}l ccc cc@{}}
\toprule
 & \multicolumn{3}{c}{HuggingFace Transformers}
 & \multicolumn{2}{c}{vLLM} \\
\cmidrule(lr){2-4}
\cmidrule(lr){5-6}
Dataset (Metric) & Full & Quest & FreeKV & Full & Ours \\
\midrule

\secrow{Long input --- Llama-3.1-8B-Instruct, LongBench~v2} \\
Overall
  & 29.62
  & 29.42 \ddb{($-$0.20)}
  & 29.03 \dd{($-$0.59)}
  & 30.23
  & 29.62 \dd{($-$0.61)} \\
Short
  & 34.44
  & 34.44 \dd{\;\;(0.00)}
  & 35.00 \dd{($+$0.56)}
  & 35.56
  & 33.89 \dd{($-$1.67)} \\
Medium
  & 28.37
  & 28.84 \dd{($+$0.47)}
  & 26.51 \dd{($-$1.86)}
  & 27.44
  & 27.44 \dd{\;\;(0.00)} \\
Long
  & 24.07
  & 22.22 \dd{($-$1.85)}
  & 24.07 \dd{\;\;(0.00)}
  & 26.85
  & 26.85 \dd{\;\;(0.00)} \\

\midrule

\secrow{Long input --- Qwen3-8B, LongBench~v2} \\
Overall
  & 32.21
  & 31.61 \dd{($-$0.60)}
  & 31.01 \dd{($-$1.20)}
  & 33.60
  & 33.20 \ddb{($-$0.40)} \\
Short
  & 36.67
  & 37.22 \dd{($+$0.55)}
  & 36.67 \dd{\;\;(0.00)}
  & 39.44
  & 38.33 \dd{($-$1.11)} \\
Medium
  & 29.30
  & 26.98 \dd{($-$2.32)}
  & 27.44 \dd{($-$1.86)}
  & 29.30
  & 30.70 \dd{($+$1.40)} \\
Long
  & 30.56
  & 31.48 \dd{($+$0.92)}
  & 28.70 \dd{($-$1.86)}
  & 32.41
  & 29.63 \dd{($-$2.78)} \\

\midrule

\secrow{Long output --- Qwen3-8B, reasoning} \\
\dm{Overall}{(pass@$k$)}
  & 81.41
  & 78.86 \dd{($-$2.56)}
  & 77.91 \dd{($-$3.50)}
  & 79.95
  & 78.69 \ddb{($-$1.26)} \\
\dm{}{(avg@$k$)}
  & 69.48
  & 66.65 \dd{($-$2.83)}
  & 66.85 \dd{($-$2.63)}
  & 67.21
  & 67.71 \ddb{($+$0.50)} \\

\dm{AIME24}{(pass@8)}
  & 86.67
  & 86.67 \dd{\;\;(0.00)}
  & 80.00 \dd{($-$6.67)}
  & 83.33
  & 86.67 \dd{($+$3.34)} \\
\dm{}{(avg@8)}
  & 77.50
  & 75.42 \dd{($-$2.08)}
  & 72.08 \dd{($-$5.42)}
  & 75.83
  & 74.58 \dd{($-$1.25)} \\

\dm{AIME25}{(pass@8)}
  & 83.33
  & 76.67 \dd{($-$6.66)}
  & 80.00 \dd{($-$3.33)}
  & 83.33
  & 76.67 \dd{($-$6.66)} \\
\dm{}{(avg@8)}
  & 70.83
  & 64.17 \dd{($-$6.66)}
  & 68.75 \dd{($-$2.08)}
  & 67.50
  & 69.58 \dd{($+$2.08)} \\

\dm{GPQA-D}{(pass@4)}
  & 74.24
  & 73.23 \dd{($-$1.01)}
  & 73.74 \dd{($-$0.50)}
  & 73.20
  & 72.73 \dd{($-$0.47)} \\
\dm{}{(avg@4)}
  & 60.10
  & 60.36 \dd{($+$0.26)}
  & 59.72 \dd{($-$0.38)}
  & 58.30
  & 58.96 \dd{($+$0.66)} \\

\bottomrule
\end{tabular}%
}
\end{table}

%% file: tex/05_related_work.tex
\section{Related Work}

KV retrieval systems retain access to the full context while moving most KV state out of HBM and staging a query-dependent working set on the GPU \cite{hisparse,spin,kvdrive,shadowkv,chen2024arkvale,chen2026retroinfer}.
ArkVale, RetroInfer, HiSparse, and SPIN place nonresident KV in CPU memory; ShadowKV instead retains low-rank keys on the GPU and offloads values, while KVDrive extends to SSD.
Because the selected nonresident entries must arrive before attention can proceed, on-demand retrieval adds latency to the decode critical path.
Existing systems reduce this cost through small retrieval budgets and HBM caching, as in HiSparse and ShadowKV \cite{hisparse,shadowkv}, or by pipelining selection and retrieval across micro-batches, as in KVDrive \cite{kvdrive}.

KV prefetch systems identify future active blocks early enough to overlap their transfer with ongoing computation.
Layer-ahead designs include InfiniGen, which rehearses the next layer from its predecessor, and HiSparse, which prefetches shared-index layers after their anchor \cite{infinigen,hisparse}.
Next-step designs include FreeKV, which reuses the previous step's selection, and SpeCache, which uses a speculative token to prefetch before the next decoding step \cite{freekv,specache}.
Learned predictors include FlashMemory-DeepSeek-V4's neural memory indexer and SparDA's per-layer Forecast projections \cite{dsv4LSA,sparda}; both require model-specific predictor training.
ECHO instead exploits the indexer already present in native sparse-attention models and does not train a separate prefetcher, although it remains tied to those model architectures \cite{ECHO-NSA}.
Training-free runtime signals avoid predictor training but often sacrifice prediction accuracy or require corrective on-demand retrieval.

Across both groups, existing designs lack a production-grade cross-tier memory manager that coordinates sparse KV placement and transfer across GPU HBM, host DRAM, and remote memory, limiting their applicability to high-throughput disaggregated serving.

%% file: tex/06_conclusion.tex
\section{Conclusion}

We presented \sys, a memory-centric LLM serving system that keeps the full KV
cache off the GPU across multiple nodes and stages only the blocks each decode step needs into HBM.
\sys proposed the accurate and general prefetch signal based on speculative decoding, and then extend KV cache pool with decode, prefill, and remote nodes based on tiered KV cache management with non-blocking asychronous prefetching considering the off-GPU bandwidth.
In evaluation, \sys preserves accuracy close to full attention on reasoning and long-context benchmarks under a constrained KV budget, while substantially improving decode throughput for both reasoning and large-batch, long-context workloads. Under PD disaggregation, it maintains high throughput while achieving high running batch size and low TTFT.

\begin{acks}
We thank Hari Govind V K, Kirill Kalinin,
Thomas Karagiannis, Xingbo Wu, Paolo Costa, Hitesh Ballani, Jacob Nelson, Dan Ports, and other colleagues at Microsoft Research for their support, valuable feedback and discussions.
We also thank Taowen Liu, Ebby Samson, Yuanxin Wei, Zixi Zhang, Zhiwen Mo and at Imperial College London for their insightful advice on this project.
\end{acks}

%% file: tex/08_appendix_fetch_cap_longbench.tex
\section{Fetch Cap on a Long-Input Workload}
\label{sec:appendix-eviction-longbench}

\Cref{tab:eviction} sweeps the per-step fetch cap on AIME24, where each request
decodes thousands of tokens and throughput is bounded by what crosses PCIe per
step. \Cref{tab:eviction-longbench} repeats that sweep on the LongBench~v2 short
split, where each request reads a long prompt but generates only a short answer.

The cap costs far less here. Every sparse configuration runs at roughly
$2.1$--$2.2\times$ dense, and loosening the cap from 0.01 to fetching everything
gives up only 8\% of that throughput (306.7 to 280.9\,tok/s), against the
$2.6\times$ collapse on AIME24. With few decode steps per request, the per-step
fetch the cap governs is a small share of the run.

Accuracy is flat across the sweep, spanning 1.7 points with no trend, and our
default ratio of 0.05 matches dense exactly on both the overall score and the
easy/hard split. The short split holds 180 questions, so one question moves the
score by 0.56 points and these gaps are a few questions wide: the reading is that
the cap costs no accuracy on this workload, not that any row is better than
another.

\begin{table}[t]
\caption{Fetch-cap ablation on the LongBench~v2 short split (Qwen3-8B, LRU
eviction). \textbf{Bold} marks our default operating point.}
\label{tab:eviction-longbench}
\centering
\small
\setlength{\tabcolsep}{5pt}
\begin{tabular}{@{}lrcc@{}}
\toprule
Fetch Ratio & TPS & Accuracy & Easy / Hard \\
\midrule
Dense          & 136.5          & 44.44          & 54.24 / 39.67 \\
\midrule
0.01           & 306.7          & 42.78          & 49.15 / 39.67 \\
0.02           & 306.5          & 41.67          & 49.15 / 38.02 \\
\textbf{0.05}  & \textbf{302.6} & \textbf{44.44} & \textbf{54.24 / 39.67} \\
0.10           & 289.0          & 43.33          & 52.54 / 38.84 \\
0.20           & 287.2          & 43.33          & 52.54 / 38.84 \\
\midrule
Fetch all      & 280.9          & 42.78          & 49.15 / 39.67 \\
\bottomrule
\end{tabular}
\end{table}

%% file: tex/11_appendix_alpha.tex
\section{Choosing the Radius Weight $\alpha$}
\label{sec:appendix-alpha}

The block score of \Cref{eq:score} weights the radius term by $\alpha$: $\alpha{=}1$ is Quest's upper bound, $\alpha{=}0$ ranks blocks by their center alone, and a negative $\alpha$ penalizes spread.
\Cref{tab:alpha} sweeps $\alpha$ from $-1$ to $1$ on Qwen3-8B at $K{=}128$ with the default fetch ratio of 0.05, on the three reasoning benchmarks and the LongBench~v2 short split.

Penalizing spread is a mistake: at $\alpha{=}{-1}$ the weighted average falls to 30.3, and the generations balloon to about $35$K tokens on AIME24 as the model loses the context it needs and rambles.
From $\alpha{=}0.25$ upward the score is stable, with the weighted average within 52.5--53.3 across the range.
Within that range $\alpha{=}0.5$ is the best overall (53.31) and the best on AIME25 and GPQA-Diamond; the full upper bound ($\alpha{=}1$) loses 5 points on AIME25, and the center alone ($\alpha{=}0$) loses 1--4 points on both AIME sets.
The long-input split is insensitive to $\alpha$, moving by at most 1.8 points with no trend.
We therefore use $\alpha{=}0.5$ throughout.

\begin{table}[t]
\caption{Radius weight $\alpha$ on Qwen3-8B ($K{=}128$, $f_{HBM}=0.05$).
avg@8 on AIME24 and AIME25, avg@4 on GPQA-Diamond and LongBench~v2 (short split); Avg weights the four by question count.
\textbf{Bold} marks our default.}
\label{tab:alpha}
\centering
\small
\setlength{\tabcolsep}{5pt}
\begin{tabular}{@{}rrrrrr@{}}
\toprule
$\alpha$ & AIME24 & AIME25 & GPQA-D & LBv2 & Avg \\
\midrule
$-1.00$       & 20.00 & 15.42 & 24.24 & 41.11 & 30.28 \\
$-0.75$       & 46.25 & 25.83 & 46.72 & 38.33 & 41.81 \\
$-0.50$       & 65.00 & 49.58 & 55.56 & 40.14 & 49.46 \\
$-0.25$       & 73.75 & 62.08 & 58.33 & 40.56 & 52.34 \\
$0.00$        & 73.33 & 65.83 & 57.07 & 41.67 & 52.45 \\
$0.25$        & 75.42 & 67.92 & 57.58 & 40.56 & 52.52 \\
$\mathbf{0.50}$ & \textbf{74.58} & \textbf{69.58} & \textbf{58.96} & \textbf{40.83} & \textbf{53.31} \\
$0.75$        & 75.42 & 66.67 & 58.71 & 40.97 & 53.11 \\
$1.00$        & 75.00 & 64.58 & 58.96 & 39.86 & 52.59 \\
\bottomrule
\end{tabular}
\end{table}

%% file: tex/12_appendix_kv_prediction.tex
\section{KV-block prediction comparison}
\label{sec:appendix-previous-vs-draft-token}

\begin{figure}[t]
	\centering
	\includegraphics[width=\columnwidth]{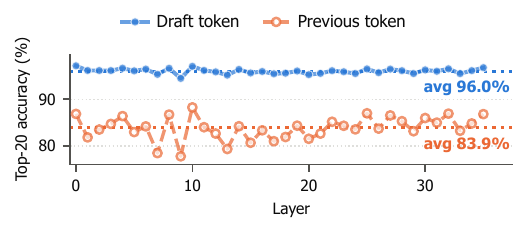}
	\caption{Top-20 KV-block prediction accuracy across model layers using the previous token versus a lookahead token.}
	\label{fig:topk_accuracy}
\end{figure}

Training-free KV prefetch designs trade accuracy for portability by reusing an inference-time signal, most commonly the current step's query as a proxy for the next \cite{freekv,specache,clo2025}. 
\Cref{fig:topk_accuracy} shows the cost of that trade: the previous-token proxy recovers the true top-20 blocks unreliably, and its accuracy varies widely across layers, so a system built on it either misses important blocks or must widen its budget to compensate.

\newpage

\section{Draft-token-driven attention prediction}
\label{sec:appendix-draft-token-prediction}

\begin{figure}[t]
    \centering
    \includegraphics[width=\linewidth]{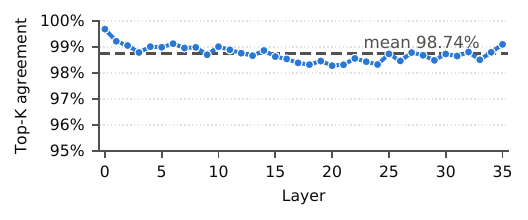}
    \caption{Per-layer agreement between the top-\ensuremath{K} set predicted by the propagated draft query and the exact set of the true next-token query.
    The profile uses Qwen3-8B on GSM8K with Tengyunw/qwen3\_8b\_eagle3 as the EAGLE-3 draft model.}
    \label{fig:topk-agreement}
\end{figure}

As observed in prior work \cite{freekv}, 
adjacent decoding steps exhibit strong temporal locality in block importance, 
so the normal token's resident blocks are likely to contain the next step's top-$K$ set.
We observe that by directly propagating the draft token over the normal token's sparse KV.
As shown in \Cref{fig:topk-agreement}, the top-$K$ set predicted by the propagated draft query agrees with that of the true next-token query above $98.2\%$ in every layer, $98.74\%$ on average.

%% file: tex/13_appendix_pd_250gib.tex
\section{PD Serving under a 250\,GiB Host Budget}
\label{sec:appendix-pd-250}

\begin{figure}[t]
\centering
\includegraphics[width=\columnwidth]{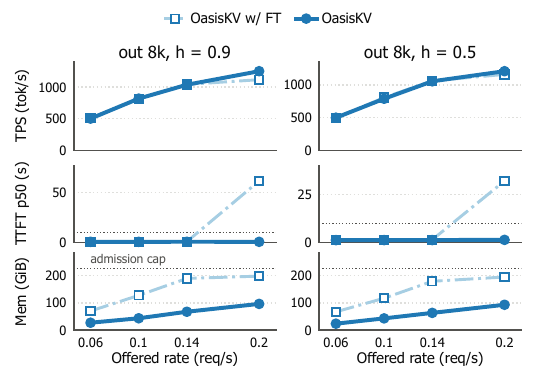}
\caption{PD-disaggregated Qwen3-8B serving with 32K-token prompts and 8K-token outputs under 250\,GiB host budgets.
Rows report throughput, median TTFT, and average decode-host memory usage; the dotted line marks the decode-host admission cap.}
\label{fig:pdserving-250}
\end{figure}

\Cref{fig:pdserving-250} repeats the 8K-output columns of \Cref{fig:pdserving} with the prefill and decode host budgets doubled to 250\,GiB each; the remote tier, the prompts, and the offered rates are unchanged.
vLLM is omitted because its dense decode is bounded by GPU KV-cache capacity rather than by the host budget (\Cref{fig:pdserving}).

At 250\,GiB the decode-host capacity no longer binds \sys{} w/ FT within the swept rates: its admission cap rises from 21 to 42 requests, above the 38--39 requests that 0.2\,req/s keeps active, so both systems run against the prefill worker's ceiling and their throughputs converge.
At 0.2\,req/s, \sys{} reaches 1{,}251 and 1{,}207\,tok/s for $h{=}0.9$ and $h{=}0.5$ against 1{,}117 and 1{,}158\,tok/s for \sys{} w/ FT, and the two are within $1.04\times$ and $1.01\times$ averaged over the four rates.

The separation moves to latency and memory.
\sys{} w/ FT ships each prompt's full KV before decoding starts and queues once its cap fills: at 0.2\,req/s its median TTFT is 62\,s ($h{=}0.9$) and 32\,s ($h{=}0.5$), against 0.8 and 1.4\,s for \sys{}, which sustains 48--49 active requests with residency filled on demand.
At that rate \sys{} holds 93--96\,GiB of decode-host memory against 194--197\,GiB for \sys{} w/ FT, and 56--59\,GiB against 139--146\,GiB averaged over the sweep.
Doubling the budget therefore buys \sys{} w/ FT throughput parity but not \sys{}'s latency: the capacity that full transfer must reserve up front, \sys{} spends only as the look-ahead selection demands it.